\documentclass[epj-ND]{svjour}
\usepackage[tbtags]{amsmath}
\usepackage{amssymb}
\usepackage{graphicx}

\usepackage{txfonts}

\begin{document}

\title{Progress in nuclear data for accelerator applications in Europe}

\author{F.Goldenbaum    \inst{1}
\fnmsep \thanks{Presenting author, \email{f.goldenbaum@fz-juelich.de}}
}

\institute{Institut f\"ur Kernphysik, Forschungszentrum J\"ulich GmbH, D-52428
J\"ulich, Germany
}

\abstract{
This contribution essentially will be divided into two parts: First, a brief
overview on topical accelerator applications in Europe, a selection of the 
European 6th
framework accelerator and ADS programs will be given, second the
emphasis will be put on selected nuclear data required for designing 
facilities planned or even under construction. In this second part the progress 
on
nuclear data in the EU FP6 Integrated Project IP-EUROTRANS (specifically 
NUDATRA) is summarized.
For proton-induced reactions in the
energy range of 200-2500~MeV experimental data and model comparisons are
shown on total and double differential production cross sections of H- and 
He-isotopes and intermediate
mass fragments.
}

\maketitle

\section{Introduction}
\label{intro}
The spectrum of nuclear data needed for accelerator applications
is extensive, versatile and complex. The complexity of this vivid field is
reflected in the large number of European (and world wide) projects currently
supported in the 6th framework programs of the EU. First, few of them will be 
briefly
presented in this contribution in order to give a coarse overview of ongoing
accelerator application actions in Europe. To a large extend community-research
infrastructure activities and integrated projects of FP6 EU-programs as
e.g.~CARE (Coordinated Accelerator Research for Europe)\cite{care}, FAIR 
(Facility for
Antiproton and Heavy Ion Research)\cite{fair}, EU-LIFE (Light Ion facility 
Europe) or IP
EUROTRANS (European Research Programme for the Transmutation of high level
nuclear waste in an accelerator driven system)\cite{eurotrans} contain domains
(e.g.~NUDATRA--Nuclear Data for Transmutation) devoted to compiling and
providing nuclear data indispensable for realization of large scale facilities.

In the framework of designing such facilities as for example subcritical
assemblies, intense pulsed spallation neutron sources (SNS \cite{spa-sns}, JSNS 
\cite{nag99}, ESSI \cite{ess02_iii}),
antiproton facilities (FAIR), accelerator-driven nuclear reactors, nuclear waste
transmutation plants, energy amplifier systems, proton drivers for future
neutrino factories (BENE--Beams for European Neutrino Experiments) and also for
the application for radioactive beams, improved accuracy nuclear data are of
great importance not only for shielding layouts, but also for estimation of
damage in target and structural materials. Under irradiation the structural
damages of materials used in construction of the facilities manifest themselves
as atomic displacements, radiotoxicity, chemical corrosion and embrittlement.
The expected radiation-induced damage of materials employed is mainly due to
\begin{itemize}
\item helium gas production
\item elastic scattering of neutrons, charged particles, and in particular
intermediate mass fragments and heavy residuals produced in spallation
reactions
\end{itemize}
Compared to the extensively studied radiation
damage in fission reactors the damage in spallation neutron sources is
characterized by an about 500 times larger ratio of produced helium gas per atom
to displacements per atom (He/dpa) in materials which are directly exposed to
the incident proton beam.
The second part of this contribution will focus on the latest progress on
nuclear data achieved in the NUDATRA domain of the IP EUROTRANS. The objective
of this project is to provide reliable and comprehensive experimental data
serving as benchmarks for code development and validation in the 200-2000 MeV
energy range.  To scrutinize several of such codes and to calculate as reliably
as possible quantities related to high energy reactions, hadronic interaction
lengths, reaction cross sections, average particle multiplicities, particle
multiplicity and double differential energy distributions need to be
investigated.  In this context the latest results of crutial experiments
performed at GSI and COSY essentially on helium and intermediate mass production
will be presented and compared to model predictions.

\section{European accelerator driven projects}
The number of small and medium sized accelerator driven projects in Europe is 
quite large. 
New ideas and initiatives are emerging and certainly worth being mentioned, 
however let us here
restrict the discussion to probably the most important forum consulted by the 
European Commission for
decision taking on supporting future facilities in Europe called ESFRI (European
Strategy Forum on Research Infrastructures) \cite{esfri06}.

\subsection{ESFRI}
The ESFRI Roadmap identifies new Research Infrastructure (RI) of European
interest corresponding to the long term needs of the European research
communities, covering all scientific areas, regardless of possible
location. Potential new RI (or major upgrade) identified are likely to be
realized in the next 10 to 20 years. Therefore they may have different degrees
of maturity. The ESFRI roadmap is an on-going process; therefore this roadmap
will be periodically updated. The first revision of roadmap will already start
early 2007. Following a request from the European Commission, ESFRI decided to
compile a list of opportunities in order to assist the Commission in the
preparation of its proposal for the Seventh Framework programme (FP7).
In autumn 2006 ESFRI agreed on a first list of 35 mature proposals for new (or
major upgrade of) facilities of European interest covering seven key
research areas (environmental sciences; energy; materials sciences;
astrophysics, astronomy, particle and nuclear physics; biomedical and life
sciences; social science and the humanities; computation, data treatment)
Those ones listed in the 2006 report \cite{esfri06} and related to accelerator 
(and reactor) applications are itemized in the following:\\
Energy:
\begin{itemize}
\item IFMIF int. fusion materials irradiation facility (10MW high flux n-source)
\item JHR high flux research reactor for fission reactors materials testing
\end{itemize}
Material Science:
\begin{itemize}
\item ELI extreme light intensity attosecond pulse laser
\item ESFR upgrade of the European synchrotron radiation facility (in 7
years)
\item ESS-I Eur.~spallation neutron source for n-spectroscopy
\item Eur.~XFEL hard X-ray Free electron laser in Hamburg
\item ILL 20/20 upgrade of the European neutron spectroscopy facility (2
phases)
\item IRUVX-FEL infrared to soft X-ray free electron lasers (in 5 user
facilities)
\end{itemize}
Astro-, nuclear and particle physics:
\begin{itemize}
\item FAIR Facility for antiproton and ion research
\item SPIRAL2 production and study of rare isotope
\item LHC large hadron collider at CERN
\end{itemize}

\subsection{Accelerator R\&D projects (EU)}
The European Commission provides a variety of instruments (I3 -Integrated
Infrastructure Initiatives, DS -Design Studies, Network Activities, NEST -New
and Emerging Science and Technology) to strengthen and financially support the
high energy physics European community in order to play a leadership role in i)
the improvement of existing accelerators ii) the development of new 
accelerators. To name a few such instruments there are e.g.:
\begin{itemize}
\item CARE Coordinated Accelerator Research in Europe, (I3)
\item BENE Beams for European Neutrino Experiments (part of CARE, Network 
Activity) 
\item EUROTeV European Design study Towards a global TeV linear Collider (plays
a major role for the ILC (int.linear collider), (Design Study)
\item EURISOL European Isotope Separation Online Radioactive Ion Beam facility, 
(Design Study)
\item EUROLEAP European laser electron controlled acceleration in plasmas to
GeV energy range, (NEST)
\end{itemize}
The duration of such projects is 3-5 years and generally about roughly 1/3 of
the total costs is covered by EU contributions.

\subsection{Multi MW-target projects}
\label{mw-target}
Doubtless and in common for any high intensity accelerator driven MW
target project nuclear data are indispensable for
\begin{itemize}
\item performance optimization (choice of material, geometry, secondary particle
production)
\item life time assessment (aging, material damage as dpa, gas production, 
embrittlement in window and target
container, corrosion,  composition modifications)
\item radioprotection (activation, radioactive inventory, short-lived residue
activity, shielding high-energy neutrons)
\item waste management (long-lived residue radiotoxicity)
\end{itemize}
Very briefly some large scale facilities currently planned or under construction 
in Europe facing during their design 
these kind of issues will be adumbrated in the following:
\subsubsection{FAIR}
The FAIR (Facility for Antiproton and Ion Research) \cite{fair} has been 
presented as large 
scale international accelerator facility of the next generation capable of 
producing primary beams of protons (up
to 30~GeV, $2.5\times 10^{13}/s$), heavy ions (up to Uranium, up to 25~GeV/u,
$10^{10}/s$) and secondary beams of radioactive ions (up to 2~GeV/u) and
antiprotons (up to 30~GeV, up to $7\times 10^{10}/h$). For the antiproton
production it is planned to bunch compress the 29~GeV protons on the production 
target to 50~ns. FAIR builds on the experience and technological developments 
already made at the existing GSI facility, and incorporates new technological 
concepts. At its heart is a double ring facility with a circumference of of 1100 
meters. A system of cooler-storage rings for effective beam cooling at high 
energies and various experimental halls will be connected to the facility. The 
new facility will be organized as a European/international research centre.
\subsubsection{Neutrino factories}
Intense neutrino factories as currently lively discussed in the EU CARE 
integrated infrastructure initiative \cite{care} and the BENE network activity
\cite{bene} employ very powerful MW proton drivers and produce finally neutrinos
from $\pi$ and $\mu$ decay, respectively. These kind of facilities aim at
$\nu$-production rates of up to $10^{21}$ useful $\mu$-decays/year.

\begin{figure}[h]
\centering
\resizebox{0.95\columnwidth}{!}{%
   \includegraphics{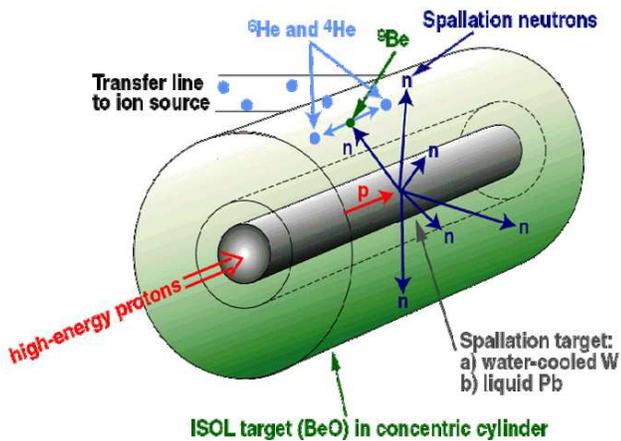}
}
\caption{$^6He$-production by converter technology using spallation neutrons.}
\label{fig:he4he6}
\end{figure}

An alternative scenario of producing intense $\nu$-beams is currently proposed 
by the decay of stored beta-active emitters instead of a $\mu$:
\begin{eqnarray}
 ^6_2He       &\rightarrow & ^6_3Li+e^-+\overline{\nu}_e \qquad 
T_{1/2}=0.8s
\label{eq:he} \\
 ^{18}_{10}Ne &\rightarrow & ^{18}_9Li+e^++\nu_e \qquad T_{1/2}=1.7s
 \label{eq:ne}
\end{eqnarray}
As shown in Fig.\ref{fig:he4he6} the He-production could be performed by 
converter technology using spallation
neutrons from water cooled W or liquid Pb on BeO concentric cylinders via the
reaction $n+^9Be\rightarrow^6He+^4He$. A nominal production rate of $5\times
10^{13}$ ions/s can be achieved. 
For the $^{18}Ne$-production a spallation of
close-by target nuclides $^{18}Ne$ from MgO $^{24}_{12}Mg(p,p_3n_4)^{18}_{10}Ne$
is assumed to be a realistic reaction--the beam hitting directly the oxide
target.However for a few GeV proton beam and 200~kW dc power the estimated
production rate will be more than one order of magnitude too low. Novel
production scenarios will be required. For the production of the specific
ions, usually an isotope separation (e.g.ISOL) is following the intense proton 
driver. These ions are accelerated to final energies and stored in large (km) 
decay rings at high $\gamma$ ($^6He: \gamma\sim150, \quad ^{18}Ne:\gamma\sim60$)
for providing neutrino sources to experiments. Post accelerating the parent ions
to relativistic $\gamma_{\textrm{max}}$ has the advantage of boosted neutrino
energy spectra and forward focussing of neutrinos: $\theta\le1/\gamma$. Two (or
more) different parent ions can be used for $\nu$ and $\overline\nu$-beams on
$\beta^+$ and $\beta^-$-decay. The physics applications of beta beams are
primarily neutrino oscillations $\nu_e \leftrightarrow\nu_\mu$ (in particularly
the single flavor decays as of eqs.\ref{eq:he} and \ref{eq:ne}) and CP violation 
studies, but also measurements on cross sections of $\nu$-nucleus interactions. 
By far however not all technical issues are addressed and solved yet.

\subsubsection{Neutron spallation source}
As a successor of the original 2003 ESS feasibility study resulting in a 
comprehensive technical report \cite{ess02_iii}, the European Spallation Source 
Initiative (ESS-I) was established as an association in which the users of 
neutron facilities, and regional consortia interested in hosting a new facility 
collaborate to prepare the case and the planning for a next generation neutron 
spallation source in Europe. The objectives of the ESS-I are to:
\begin{itemize}
\item gather all those interested in building at the earliest possible 
occasion a next generation top tier neutron source in Europe
\item stimulate, coordinate where necessary and possibly engage in all 
activities that need to be carried out before a decision to recommence baseline 
engineering, construction planning and detailed site assessment can be taken
\item act as the interlocutor to national governments and funding agencies, 
the EU and the ESF in matters regarding the initiative
\item have one European counterpart to SNS and the JSNS as part of J-PARC
\end{itemize}
Related to the project laid down in the ESFRI roadmap, i.e. the 5~MW long pulse 
target station, several countries/governments have declared to take an active 
part in the "preparatory phase" project of ESS within EU FP7 and shown interest 
in hosting the large scale facility. However appropriate conditions for 
international co-financing are not set up so far. 

A baseline design phase resulted in a series of Volumes \cite{ess02_iii} by the 
end of 2003 addressing the technical challenges that had been identified. As 
concerns the need for nuclear data, of crutial importance of any revised layout 
will be the optimization of the target/moderator/reflector assembly (geometry, 
material,...) and the maximization of neutron flux. Carefully estimated also 
have to be the production rates in GeV p-induced spallation on heavy targets and 
the activities of target and structural materials during and after many years of 
full power operation. 
\begin{figure}[h]
\centering
\resizebox{0.95\columnwidth}{!}{%
   \includegraphics{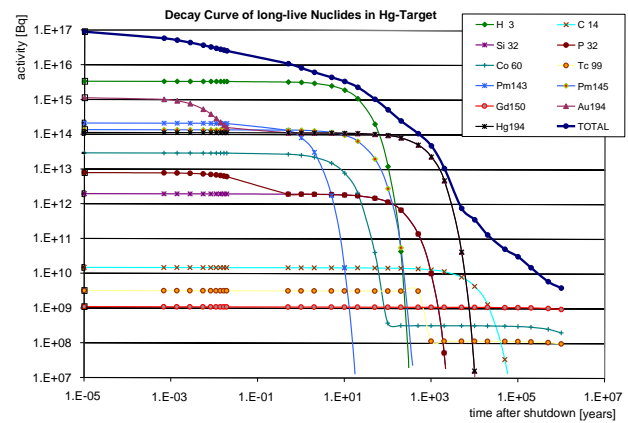}
}
\caption{Decay curves of long lived nuclides in Hg-target which determine the 
activity during operation and after 30years of 5~MW operation.}
\label{fig:nuclide-decay2}
\end{figure}
Figure \ref{fig:nuclide-decay2} shows as an example for a Hg target the typical 
decay curves of individual nuclides during operation and after 30years of 5~MW 
operation as for ESS conditions. Of largest abundance is here the tritium 
activity up to t=10 years as well as the $^{194}Hg$ and $^{194}Au$ activity 
dominating up to 1000 years and representing at the same time as long lived 
isotopes the nuclides of largest concern. 

\section{Nuclear data EU-projects}
Already in the fifth framework program, projects aiming at new nuclear data 
measurements and evaluations as e.g. HINDAS (High and Intermediate energy 
Nuclear Data for Accelerator-driven Systems) were supported by the EU. Within 
the three years program HINDAS provided a wealth of new nuclear reaction cross 
sections in the energy range between 20 MeV and 2~GeV on three elements of 
crutial importance for ADS systems: Pb as a target element, U as an actinide and 
Fe as a shielding element. As a natural extension or successor of the FP5 HINDAS 
project, in FP6, the project EUROTRANS (EUROpean Research Programme for the 
TRANSmutation of high level nuclear waste in an accelerator driven system) has 
been launched. Due to the lack of progress in implementing the radioactive waste 
disposal in geological repositories, the Partitioning and Transmutation (P\&T) 
strategy has been put forward.  
requirements. In addition, P\&T both will contribute to the sustainability of 
nuclear energy in those countries that pursue this source of energy, and will 
assist in combating global warming.
In the Euratom Sixth Framework Program (FP6), one specific targeted research 
project on the study of the impact of P\&T (RED-IMPACT) and two integrated 
projects: (i) EUROPART (partitioning) and (ii) EUROTRANS (transmutation) have 
been launched. EUROTRANS benefits not only from the technological developments 
and scientific progress achieved in Europe during FP5, but also from worldwide 
co-operation (OECD/NEA, IAEA, USA, Japan, ISTC etc.) in the field of P\&T. The 
FP5 R\&D results in the areas of preliminary design studies of an experimental 
ADS, low-power coupling of an accelerator to the Masurca zero power reactor made 
sub-critical for the purpose of the MUSE programme, studies on transmutation 
fuels, heavy liquid metal technology and high-energy nuclear data are 
available as a basis for the advancement of this Integrated Project (IP). The 
EUROTRANS project is structured into one management and five technical Domains 
(DM) that are further subdivided into work packages and tasks. 

\section{Selected nuclear data (NUDATRA)}
In the following the emphasis will be on EUROTRANS - DM5 NUDATRA (Nuclear data 
for transmutation of nuclear waste). The goal of this domain is to improve 
nuclear data evaluated files and models which involves sensitivity analysis and 
validation of simulation tools, low and intermediate energy nuclear data 
measurements, nuclear data libraries evaluation at low and medium energies, and 
high energy experiments and modeling. In the following the focus is on NUDATRA 
WP5.4---High energy experiments and modeling. This workpackage aims at the 
investigation of: 
\begin{itemize}
\item pA (spallation) reactions in the GeV regime
\item data measured from exclusive experiments for testing, validating and 
developing theoretical models
\item double differential cross sections (DDXS) $d\sigma/dEd\Omega$ of light 
charged particles (LCP=p,d,t,$^3$He,$^4$He,...) and intermediate mass fragments 
(IMFs, $Z\le 16$) in spallation and fragmentation p-induced reactions 
(0.1-2.5~GeV, C to Au)
\item reaction mechanism of pN reactions in terms of time scales, simultaneous 
or sequential emission of IMFs, origin of pre-equilibrium and evaporation 
processes
\end{itemize}

\subsection{Light charged particle and IMF production}
\begin{figure}[h]
\centering
\resizebox{0.95\columnwidth}{!}{%
   \includegraphics{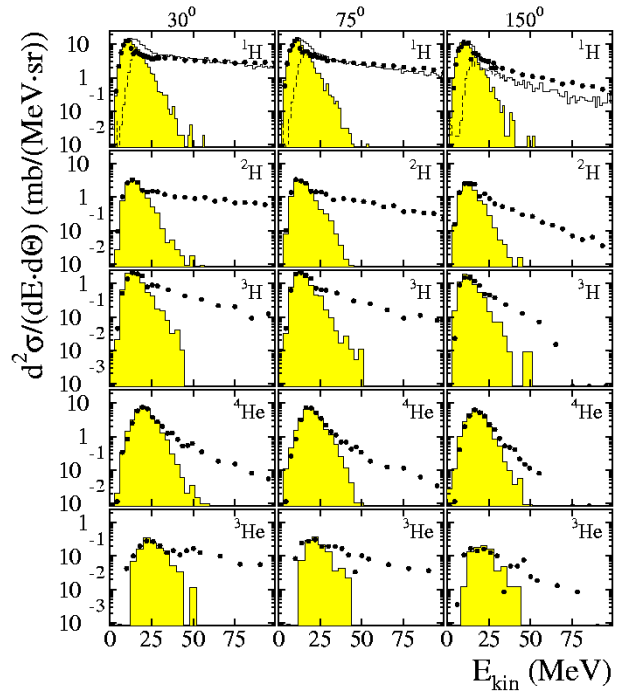}
}
\caption{Energy spectra of $^{1,2,3}$H and $^{3,4}$He for 1.2~GeV p+Ta. dots: 
exp.~data, shaded hist.: calculated evap.~spectra, dashed hist.: pre-equilibrium 
protons as calculated by INCL2.0; \protect\cite{her06}}
\label{fig:fig7a}
\end{figure}

As an example for light charged particle production, double differential energy
spectra of $^{1,2,3}$H and $^{3,4}$He ejectiles following 1.2~GeV p-induced 
reactions on Ta target as measured by NESSI collaboration at COSY-J\"ulich is 
shown for different angles in respect to the incident proton in 
fig.\ref{fig:fig7a}. 
The experimental data clearly feature two components, an evaporation component 
dominant for all angles and at low kinetic energies and a high energy component 
all the more pronounced the smaller the angle of the ejectile in respect to the 
incident proton is. Here \cite{her06} for the theoretical description the 
INCL2.0 \cite{cug97} intranuclear cascade code is coupled to the evaporation 
code GEMINI \cite{cha88}. Only for protons both components can be well 
described. Due to the lack of composite particle emission in the early phase of 
the reaction in the INCL2.0 model, the high energy tails of the spectra for 
d,t,$^{3,4}$He are not described by the calculations. The shape of the 
calculated evaporation component (shaded yellow histogram in 
fig.\ref{fig:fig7a}) however is well reflected also for composite particles. 
\begin{figure}[h]
\centering
\resizebox{0.95\columnwidth}{!}{%
   \includegraphics{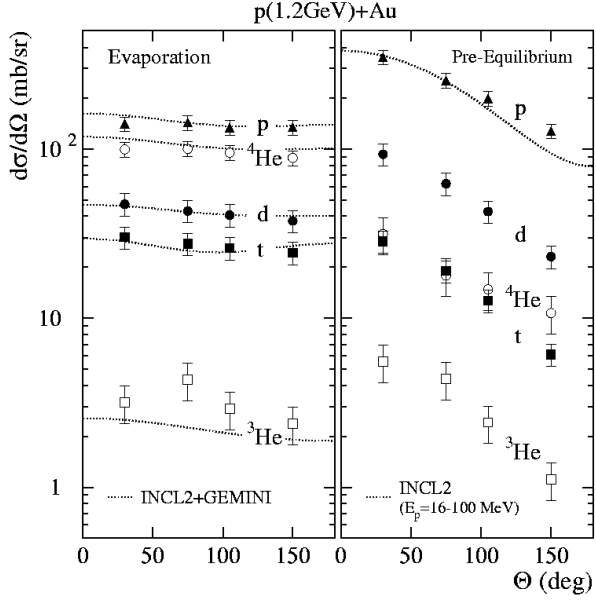}
}
\caption{Angular distributions of $^{1,2,3}$H and $^{3,4}$He for 1.2~GeV p+Au. 
symbols: exp.~data, lines calculation by INCL2.0+GEMINI; \protect\cite{her06}}
\label{fig:fig8}
\end{figure}
For 1.2~GeV p+Au in fig.\ref{fig:fig8}, the angular distribution of disentangled 
evaporation (left panel) and pre-equilibrium (right panel) components are shown. 
For all particle species the evaporation exhibits an isotopic behaviour, while 
more directly emitted particles show larger forward/backward asymmetry. Note 
that for pre-equilibrium protons the angular dependence is well described in the 
INCL2.0 model. It would be certainly worth to compare the current experimental 
data \cite{her06} with the latest version of INCL4.3 \cite{bou04} including a 
coalescence formalism allowing for the cluster emission of composite nucleons 
(d,t,$^{3,4}$He) in the early phase of the reaction. 

\begin{figure}[h]
\centering
\resizebox{0.95\columnwidth}{!}{%
   \includegraphics{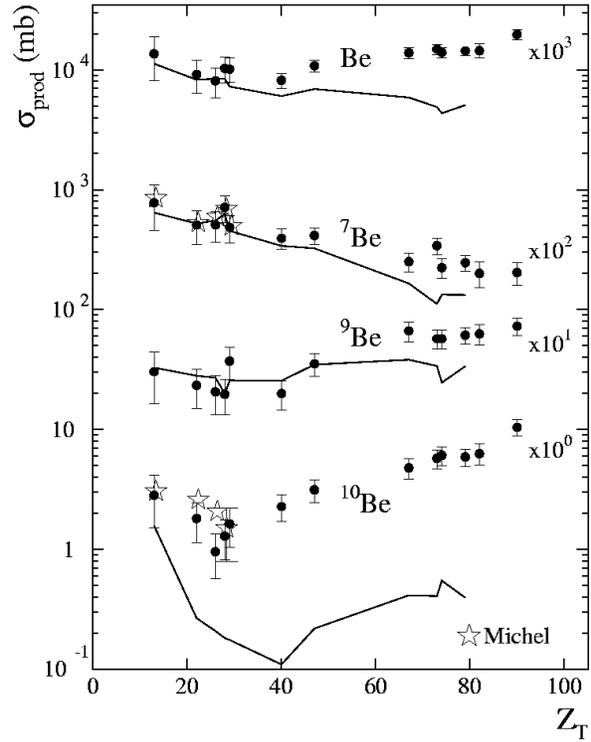}
}
\caption{Production cross section of $^{7,9,10}$Be isotopes for 1.2~GeV p+X. 
$\bullet$: NESSI \protect\cite{her06}, $\star$: R.~Michel \protect\cite{mic95} 
data, lines: calculation by INCL2.0+GEMINI}
\label{fig:fig16}
\end{figure}
The production cross sections of $^{7,9,10}$Be isotopes for 1.2~GeV protons on 
different targets (C to Au) as well as the total $\sigma_{\textrm{Be}}$ are 
shown in fig.\ref{fig:fig16}. The production of all individual isotopes does not 
strongly depend on $Z$, respectively. When looking more carefully at the energy 
spectra of IMFs (not shown in this contribution), ones more as expected the 
combination of INCL2.0+GEMINI fails to describe the high energy tails of the 
energy spectra. Nevertheless in fig.\ref{fig:fig16} the calculated angle and 
energy integrated production cross sections agree generally rather well with the 
NESSI \cite{her06,mic95} data, because the pre-equilibrium component amounts to 
the total cross section only on the percent level. The lines representing the 
model prediction are reflecting the ejectiles coming from evaporation model 
only, i.e.~GEMINI. The experimental data on $^7$Be and $^{10}$Be ejectiles 
measured for low $Z$-targets by mass spectrometry \cite{mic95} coincide with the 
systematics of the NESSI experiment \cite{her06}. In a similar presentation one 
would observe the multiplicity/production cross sections of the neutron rich He 
isotope ($^6$He) strongly increasing with atomic number $Z$ of the bombarded 
target (not explicitly shown here)--a very similar behavior as the one which is 
observed for the "neutron rich" triton. In contrast to the $^{3,4}$He isotopes, 
for $^6$He the INCL2.0+GEMINI calculations {\em overestimates} the experimental 
results of Herbach et al.\cite{her06} by approx.~30\%. 

The international collaboration PISA (Proton Induced SpAllation) \cite{pisa-web,gol05,bar03,bub04} is aiming at a quite similar physics program as NESSI, however with a completely different setup and at an internal beam location. At the internal beam of COSY the investigation of the reactions induced by protons on thin targets (50-200 $\mu$g/cm$^2$) enables us to get the cross sections without the absorption and energy loss involved with the propagation of reaction products in the material of the target. The multiple circulation of the beam in the COSY ring is used to compensate for the small reaction rate of beam-protons with the thin targets. The advantage being higher statistics and more precise information on the very tails of double differential energy spectra.
Isotope separation was done by combining the information from multi-channel-plates (time-of-flight), silicon detector telescopes and  Bragg curve spectroscopy (energy deposited inside Bragg curve detectors) allowing for the separation of following isotopes: 6Li, 7Li, 8Li - 7Be, 9Be, 10Be - 10B, 11B - 11C, 12C, 13C, 14C and 13N, 14N \cite{bub04}. Measurements of these cross sections 
are important for providing benchmark data in the GeV incident p- energy range, understanding the complex reaction mechanism itself and testing the reliability of physical models describing the fast intranuclear cascade (INC) phase as well as the subsequent statistical decay from an equilibrated or thermalized hot nucleus. As already mentioned, a particular focus is on developing new models for the description of highly energetic composite particles \cite{bou04}. 

\begin{figure}[h]
\centering
\resizebox{0.95\columnwidth}{!}{%
   \includegraphics{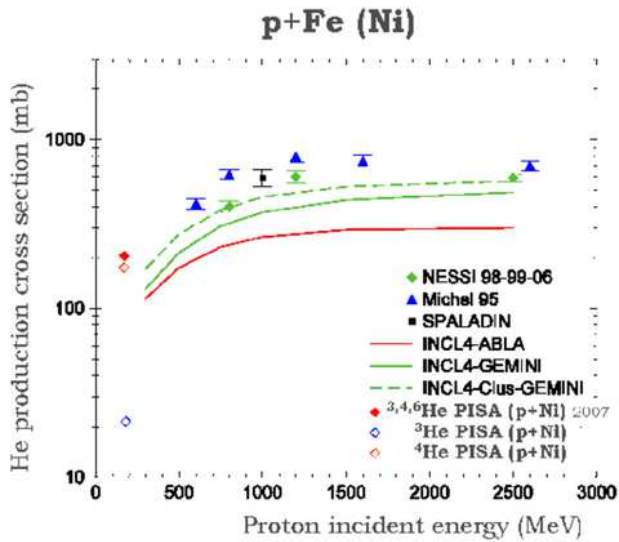}
}
\caption{Production cross section of $^{3,4,6}$He isotopes as a function of incident proton beam energy. symbols: NESSI,Hannover,SPALADIN,PISA data, curves: calculation by INCL4.3+GEMINI/ABLA}
\label{fig:he-prod}
\end{figure}
As a function of incident proton beam energy the He-production cross sections on 
Fe measured by NESSI\cite{her06}, Hannover\cite{mic95}, SPALADIN\cite{gen06}, 
and PISA\cite{pis07} are compiled in fig.~\ref{fig:he-prod}. The latest data 
points of SPALADIN at 1~GeV and PISA at 175~MeV are also included. The SPALADIN 
result obtained in inverse kinematics of Fe on p at GSI shows a value slightly 
above the NESSI data, but is definitely still smaller than the systematics of
R.~Michel et al.\cite{mic95}. The data shown here for PISA are for Ni reaction, 
but a comparison should be legitim, because Fe and Ni are very close in terms of 
atomic number. Note, that for the PISA data \cite{pis07} the cross sections for 
the individual $^{3,4,6}$He isotopes are given at 175~MeV. The Monte Carlo 
calculation getting closest to the available experimental He data is the 
INCL4-Clus-GEMINI version (dashed line in fig.~\ref{fig:he-prod}), which 
accounts-using a coalescence approach-for cluster (here composite He particles) 
emitted in the first fast phase of the reaction. The two solid lines in 
fig.~\ref{fig:he-prod} take into account only the He particles being emitted 
during the slow evaporation phase and therefore as expected the abundance of 
production cross sections is underestimated in INCL4+ABLA or INCL4+GEMNI, 
respectively. 

Of great value and particular interest are the measurements performed by mass spectrometry \cite{mic95,ley05} The authors provide excitation functions in the whole energy range of interest, however in particular for light targets typically the measured He production cross sections do not coincide. The discrepancies between the two experimental methods for light targets is not yet understood. The huge amount of data collected for proton induced reactions will be valuable for the identification of deficiencies of existing INC/evaporation codes. 

\subsection{Neutron production}
Of significant interest for a wide range of applications and fundamental
research, in particular at the crux of spallation neutron sources, transmutation 
of nuclear waste in accelerator driven systems \cite{nif01}, and shielding 
issues are also neutron production double differential cross sections in GeV
proton-induced spallation reactions. Altough generally best described by
INC+evaporation codes, neutrons are more difficult to detect than protons or
LCP. Experimental double differential neutron production spectra represent a
valueable observable also for validating new model developments or improvements
\cite{bou04,cug97b,dua07,bou02}. It is also interesting to look at neutron
multiplicities $M_n$, global properties of neutron spectra which are not easily
revealed by their inspection. An extensive overview on the observable $M_n$ for
both thin and thick (ejectiles induce secondary reactions) targets is compiled
in Refs.~\cite{bou02,fil01,let00,hil98}.

\section{Conclusion}
It has been shown, that a series of large scale accelerator facilities are currently planned or even under construction in Europe. The EU financially supports activities and provides within framework programs several instruments 
for pushing forward the realization of such facilities.  The spectrum of nuclear data needed for accelerator applications is quite broad and complex and certainly for the realization of facilities at the border of todays technology, new nuclear data are indispensable for estimates on radiation damage by displacements or embrittlement. 
Some selected aspects within the Integrated Project EUROTRANS domain NUDATRA have been presented. Few experiments have been consulted to validate models with regard to reaction cross-sections or reaction probabilities, charged particle production cross-sections and angular- and energy- distributions for GeV proton induced reactions on various thin targets. PISA experiment e.g.~has showm to be able to measure 
the products of proton--nucleus collisions with Z-identification up to at least Z=16 and isotope identification to masses up to 13-14.
A comprehensive comparison not only of light charged particles, but also of IMF particles with model predictions is in progress. Therefore the high energy experiments presented here provide an important set of benchmark data for the development and test of reliable new models capable of describing the emission of the high energy component of composite particles.

\begin{acknowledgement}
The author acknowledges gratefully the support of the European Community-Research Infrastructure Activity under FP6 "Structuring the European Research Area" programme (CARE-BENE, contract number RII3-CT-2003-506395 and HadronPhysics, contract number RII3-CT-2004-506078). The NESSI/PISA collaboration appreciates the financial support of the European Commission through the FP6 IP-EUROTRANS FI6W-CT-2004-516520.
\end{acknowledgement}

%

\end{document}